\documentclass[letterpaper, a4paper]{amsart}

\usepackage{amsmath}
\usepackage{latexsym}
\usepackage[all]{xy}
\usepackage{color}

\include{amslatex}

 \numberwithin{equation}{section}

\begin{document}
\begin{title}[On Penrose's gravitationally wave function reduction]
 {On Penrose's theory of objective gravitationally induced wave function reduction}
\end{title}
\clearpage\maketitle
\thispagestyle{empty}
\begin{center}
\author{Ricardo Gallego Torrom\'e\footnote{Email: rigato39@gmail.com}}
\end{center}
\bigskip

\begin{center}
\address{Department of Mathematics\\
Faculty of Mathematics, Natural Sciences and Information Technologies\\
University of Primorska, Koper, Slovenia}
\end{center}

\begin{abstract}
The formal structure of Penrose's gravitationally induced reduction of the wave function mechanism is analyzed. It is shown that pushing Penrose's argument forward leads to the interpretation of quantum coherence in microscopic systems as an observable signature violation of general covariance. We discuss potential avenues to avoid this conclusion, among them emergent quantum mechanics and super-determinism.
\end{abstract}

\section{Introduction}
A proper understanding for the absence of quantum superpositions in macroscopic systens is of fundamental interest in the foundations of quantum theory. One class of schemes aimed to address this question involves gravity as the origin of the reduction mechanism \cite{Diosi 1987,Diosi 1989, Penrose 1996,Penrose 2005, Penrose 2014a}. In this paper we will discuss the approach taken by Penrose, view as an attempt to preserve as maximal as possible the principles of general relativity as providing a mechanism for the absence of superposition  \cite{Penrose 1996}.
 The mechanism proposed by Penrose involves the gravitational interaction in a fundamental way, by means of the role of the principle of general covariance and the deep interrelation between the geometric properties of spacetime and the distribution of matter in it  as they appear implemented in general relativity and the non-relativistic Newton-Cartan theory.

 The principle of general relativity as used in Penrose's theory is the hypothesis that the geometry of the physical spacetime is represented by a formalism which is covariant with respect to general spacetime coordinate transformations and that a reduction of the possible transformations allowed by the theory will not simplify the formalism (see \cite{Bergmann 1976}, Chapter X). This formal principle is vindicated in general relativity for consistency with the weak equivalence principle. The standpoint that such principles (general covariance and geometric nature of gravity) should be kept at a fundamental level is the motivation of Penrose's theory \cite{Penrose 1996,Penrose 2005,Penrose 2014a}.
However, according to Penrose, one can make progress in the problem of finding the mechanism for the absence of quantum superpositions in macroscopic systems if small deviations from the principle of general covariance are allowed. Under this less restrictive condition, one can provide a general mechanism for the reduction of wave packet. The mechanism is such that for small systems the reduction will take a long time, but for large systems the reduction will act rapidly.

Despite Penrose's theory relies on fundamental principles and provides a natural mechanism for the reduction of the wave packet, we would like to highlight in this paper one problematic issue related to the proposal. We argue that when Penrose's mechanism is applied to small systems, then the existence of quantum coherence, usually presented in quantum systems, has to be read as an observable violation of general covariance. This is in conflict with the starting motivation of the argument, namely that general covariance should be kept safe at macroscopic and microscopic scales or that it should be kept safe, even if in an approximated way in the last case. In particular, we argue that the measure of the violation of general covariance proposed by Penrose is not the most adequate and that when another more appropriate measure is used, the same amount of violation of general covariance should be present for macroscopic and for microscopic scales. This obviously poses a paradox between the aims and the result of the argument.

However, if  Penrose's argument, although currently in incomplete form, is assumed essentially tenable, then there are two different minimal resolutions of the above paradox:
1. Either there  is no real quantum coherence, as has been proposed by some authors \cite{Hooft2012wavefunctionschroedingercollapse,Hooft2016} or 2. General covariance is violated at the quantum level when quantum coherence is present. Both possible resolutions imply dramatic changes in perspective in the foundations of physics, reflect the deepness and relevance of Penrose's mechanism.
\section{Short review of Penrose's theory}
Penrose's theory of gravitationally induced collapse of the wave function is not a complete theory of objective reduction of the wave function, but it illustrates a general mechanism for the reduction that involves gravity in a fundamental way, using  the less possible technical context. In Penrose's theory, a geometric Newtonian gravitational framework, Cartan formulation of Newtonian gravity, is assumed. Cartan's general covariant formulation of Newtonian gravity is useful to directly full-fill the required compatibility with the principle of general covariance.
 The adoption of Newton-Cartan theory is also justified by the type of experiments suggested by Penrose's argument that can be implemented in the Newtonian limit.

 Therefore, a brief introduction to Cartan's geometric framework is in order. For a comprehensive treatment, see \cite{Cartan1923, Cartan1924} and for a  modern presentation, \cite{MisnerThorneWheeler}, Chapter 12. The spacetime arena in Newton-Cartan theory is a smooth four dimensional manifold $M_4$.  There is defined on $M_4$ an affine, torsion-free connection $\nabla$ that determines free-fall motion as prescribed in Newtonian gravity.  The scalar field $t:M_4\to \mathbb{R}$ is called the {\it absolute time} and it determines the $1$-form $dt$ and finally, there is a three dimensional Riemannian metric $g_3$ on each section $M_3$ transverse to $dt$: thus $X\lrcorner dt=0$, for each $X\in \,TM_4$ tangent to $M_3$. The connection $\nabla$ preserves the $1$-form $dt$. The metric $g_3$ is also compatible with the connection $\nabla$. An analogue of Einstein's equations define the dynamical equations for $\nabla$ in terms of the density of matter $\rho$. The Riemannian metric $g_3(t)$, defined on each transverse space $M_3(t)$, is determined by further geometric conditions. For practical purposes, in this paper can be thought $g_3$ as a given, although Newton-Cartan establishes a complete dynamical theory for $g_3$ \cite{MisnerThorneWheeler}.

 There are two further assumptions that Penrose uses in his argument and that are worthily to mention here:
 \begin{itemize}

 \item Slightly different spacetimes $M_4(1)$, $M_4(2)$, corresponding two different configurations of the system are such that the absolute time functions $t_i:M_4(i)\to\,\mathbb{R}, \,i=1,2$ can be identified.

 \item The spacetime is stationary, that is, there exists a Killing vector of the metric $g_3$,
 \begin{align}
 \nabla_T\,g_3=0.
 \label{timelike killing vector}
 \end{align}
 \end{itemize}

Some comments are in order.
First, let us remark that the identification of the absolute coordinate $t_1\in \,M_4(1)$ with the absolute coordinate $t_2\in\,M_4(2)$ is motivated by technical reasons, since it avoids to consider several fine details in the calculations.
 Second, a timelike Killing vector on $M_4$ determines a well-defined notion of stationary state in a stationary Newton-Cartan spacetime. Regarded as a derivation of the algebra of complex functions $\mathcal{F}(M_4)$, the fact that $T$ is globally defined implies that the eigenvalue equation
 \begin{align}
 T\,\psi =\,-\imath\,\hbar\,E_\psi \,\psi
 \label{stationary state}
 \end{align}
is consistent on $M_4$.

After the short introduction of the geometric setting, let us highlight the logical steps of Penrose's argument as it appears for instance in  \cite{Penrose 1996} or in  \cite{Penrose 2005}, Chapter 30. The experiment considers the situation of superposition of two quantum states of a lump of matter, when the gravitational field of the lump is taken into consideration.
We have articulated Penrose's argument as follows:
\\
{\bf 1.} The principle of general covariance is invoked and it is shown its incompatibility with the formulation of stationary Schr\"odinger equations for superpositions when the gravitational fields of the quantum systems are taken into consideration. In particular, Penrose emphasizes why one cannot identify two diffeomorphic spacetimes in a pointwise way. Therefore, the identification of timelike killing vectors pertaining to different spacetimes is not possible and one cannot formulate the stationarity condition \eqref{stationary state} for superpositions of lumps of matter, since each lump determines its own spacetime. Each time derivative operation is determined by the corresponding timelike Killing field and each of those Killing fields lives over a different spacetime $M_4(1)$ and $M_4(2)$. Note that this obstruction is absent in usual quantum mechanical systems, where the gravitation fields of the quantum system is systematically disregarded as influencing the spacetime arena, namely, the Galilean or the Minkowski spacetime.
\\
{\bf 2.} If one insists on identifying Killing vectors of different spaces, it will be an {\it error} when doing such identification. Such an error is assumed to be also a measure of the amount of violation of general covariance. Penrose suggested a particular measure $\Delta$  and evaluated it in the framework of Cartan's formulation of Newtonian gravitational theory by identifying the corresponding $3$-vector accelerations associated with the corresponding notions of free fall.
\\
{\bf 3.} An interpretation of the error in terms of the difference between the gravitational self energies of the lumps configurations $\Delta E_G$ is developed, with the result that
$\Delta =\,\Delta E_G.$
\\
{\bf 4.} Heisenberg's energy/time uncertainty relation is applied in an analogous way as it is applied for unstable quantum systems to evaluate the lifetime $\tau$ for decay  due to an instability. In the present case,  the energy uncertainty is the gravitational self-energy of the system $\Delta\,E_G$. It is then assumed that such energy uncertainty associated to the superposition of different spacetimes is associated to an unstable system, whose lifetime is
\begin{align}
\tau\sim \,\frac{\hbar}{\Delta\,E_G}.
\label{life time}
\end{align}
\\
{\bf 5}. It is then hypothesized that the system can persists in a violation of general covariance during a time $\tau$ given by the expression \eqref{life time}.
For macroscopic systems $\hbar/\Delta\,E_G$ is a very short time. Therefore, the argument provides an universal mechanism of gravitational induced objective reduction of the wave function for macroscopic systems.

The cause for the gravitational induced objective reduction of the wave packet is a form of perturbation of the stationary states due to the difference on the gravitational self-energy associated to the different spacetimes $M_4(1)$ and $M_4(2)$. This ultimately is associated with the violation of the principle of general covariance. The existence of a difference of energy $\Delta=\,\Delta\,E_G$ makes that the system in superposition cannot be stationary and will decay to a stationary state. Apart from this general mechanism for reduction, Penrose's argument does not discuss a particular dynamics for the gravitationally induced collapse mechanism.

 Penrose's argument relies on the  following construction. The estimated error in the approximation performed in the points 2. and 3. is given in Penrose's theory by the integral
 \begin{align}
 \Delta =\,\int_{M_3(1)}\,d^3 x\,\sqrt{\det g_3}\,g_3(\vec{f}_1(t,x)-\vec{f}_2(t,x),\vec{f}_1(t,x)-\vec{f}_2(t,x)),
 \label{error in approximation}
 \end{align}
 where $\vec{f}_1(t,x)$ and $\vec{f}_2(t,x)$ are the acceleration 3-vectors of the free-fall motions of test particles for the connection determined by the lumps configurations at the positions position $1$ and $2$, but when $x$ is regarded as points of $M_3(1)$, submanifold of $M_4(1)$. $M_3(1)$ depends on the value of the absolute time parameter $t=t_0$.
 Although the measure $\int_{M_3(1)}\,d^3 x\, \sqrt{\det \,g_3}$ is well defined and invariant under transformations leaving the $1$-form $dt$ and the space submanifold $M_3(1)$ invariant, the integral \eqref{error in approximation} is an ill-defined object. This is because $\vec{f}_2(t,x)$ is not defined over $M_3(1)$ but over $M_3(2)$ and hence, the difference $\vec{f}_1(t,x)-\vec{f}_2(t,x)$ is not a well-defined geometric object. However, in Penrose's argument the integral operation \eqref{error in approximation} is understood as an indicator of an error due to an assumed violation of general covariance.
To play the role of a meaningful error estimate, the integral \eqref{error in approximation} at least must have an invariant meaning, independent of any diffeomorphism leaving the $1$-form $dt$  and each space manifold $M_3(1)$ invariants. This is indeed the case, module the issue of the problematic nature of  $g_3(\vec{f}_1(t,x)-\vec{f}_2(t,x),\vec{f}_1(t,x)-\vec{f}_2(t,x))$.

 Assuming that Penrose's argument is applicable, it can be shown that the integral \eqref{error in approximation} is related with the differences between Newtonian gravitational self-energies of the two lumps configurations $\Delta =\,\Delta E_G$ \cite{Penrose 1996},
\begin{align}
\nonumber \Delta =\,\Delta E_G =\,-4\,\pi\,G\,\int_{M_3(1)}\,\int_{M_3(1)}\,d^3x\,d^3 y \,&\sqrt{\det \,g_3(x)}\sqrt{\det\,g_3(y)}\,\cdot\\
& \cdot\frac{\left(\rho_1(x)-\rho_2(x)\right)\,\left(\rho_1(y)-\rho_2(y)\right)}{|x-y|}.
\end{align}

\section{On the application of Penrose's theory to microscopic systems}
In the following paragraphs we discuss a paradoxical consequence of Penrose's gravitationally induced reduction mechanism of the quantum state theory.
 We first remark that strictly speaking,  Penrose's argument should also be applicable to {\it small scale systems} in the following form. First,  there is no scale in Penrose's argument limiting the applicability of the error measure \eqref{error in approximation}  for the violation of the general covariance. Therefore, the same considerations as Penrose highlights for macroscopic systems, must apply to small systems. For small scale systems, quantum coherence is an extensively experimental corroborated phenomena and it is one of the fundamental concepts of quantum mechanical description of physical systems. Thus by Penrose's argument, the existence of microscopic quantum coherence seems inevitably to be interpreted as an observable violation of general covariance. Despite it is usually argued that such a violation is small, we will show that this is not the case and that indeed such violations constitute a threat that leads to a contradiction between the aims of Penrose's theory and its consequences.

 Let us first agree in that the violation of  general covariance is measured by the identification $\Delta (t)=\,\Delta E_G(t)$. $\Delta E_G(t)$ is a function of the absolute time coordinate function $t:M_4(1)\to \,\mathbb{R}$, where we have associated a dependence on time $t$ to the error and gravitational self-energies, because they formally depend on time $t$ through the integral operations $\int_{M_3(1)}$ and each $M_3(1)$ is defined at constant time $t$. For typical quantum systems, $\Delta E_G (t)$ is very small, since the associated gravitational field far from the own locations  is weak. However, due to the large time that coherence could happen for a microscopic system, $\Delta E_G(t)$  is not the best measure of violation of general covariance. For microscopic systems, coherence in energy due to superpositions of spacetimes can persist for a long interval of time $t$. In such situations and assuming that $\Delta E_G(t)$ is constant on $t$, it is the quantity $\tau\,\Delta E_G$ what appears to be a better measure for the violation of the general covariance, where here $\tau$ is the span of absolute time $t$ such that the superposition of spacetimes survives.

Let us elaborate further on the above idea. The measure of the violation of general covariance proposed by Penrose is an integral of the modulus of the difference between two vector fields in space given by \eqref{error in approximation}. Such measure cannot take into account the possible accumulative effect in the violation of general covariance that a persistent quantum coherence in a given system can have. This train of thoughts suggest that $\Delta E_G$ must be considered as an {\it error density}, while the {\it error} in the violation to general covariance should be obtained by integrating $\Delta E_G(t)$ along the spam $\tau$ of absolute time $t$ that the superposition persists. Therefore, the error in the approximations due to the violation of general covariance by a local identification of spacetimes due to superposition of lumps must be given by the expression
\begin{align}
\widetilde{\Delta} =\,\int_{M_4(1)}\,dt\wedge \,d^3x\,\sqrt{\det g_3}\,g_3(\vec{f}_1(x,t)-\vec{f}_2(x,t),\vec{f}_1(x,t)-\vec{f}_2(x,t)),
\label{covariant error in the approximation}
\end{align}
Note that with the geometric structures available in the Cartan-Newton space considered by Penrose, the natural invariant volume form in $M_4(1)$ that one can construct is $dt\wedge\,d^3 x\,\sqrt{\det\,g_3}$.  Furthermore, the volume form $dt\wedge \,d^3x\,\sqrt{\det g_3}$ is invariant under the most general diffeomorphism of $M_4(1)$ leaving the $1$-form $dt$ invariant, a property which is not shared by the form $d^3x \,\sqrt{\det g_3}$ assumed by Penrose's theory. In these sense, the measure $\widetilde{\Delta}$ is unique, supporting \eqref{covariant error in the approximation} instead than \eqref{error in approximation} as covariant measure. Furthermore, if  $g_3(\vec{f}_1(x,t)-\vec{f}_2(x,t),\vec{f}_1(x,t)-\vec{f}_2(x,t))\neq \,0$ but constant only in an interval $[0,\tau]$ and  alsewhere  zero, then
\begin{align*}
\widetilde{\Delta} & =\,\int^\tau_0 dt \,\int_{M_3(1)}\,d^3 x\,\sqrt{\det g_3}\,g_3(\vec{f}_1(t,x)-\vec{f}_2(t,x),\vec{f}_1(t,x)-\vec{f}_2(t,x)) =\,\tau\,\Delta\\
& =\,\tau\,\Delta E_G
\end{align*}
holds good, demonstrating the interpretation of $\widetilde{\Delta}=\,\tau\,\Delta E_G$ as the value of a four dimensional integral on $M_4(1)$.

 Analogously as in Penrose's argument, in order to determine the lifetime $\tau$ one applies Heisenberg energy/time uncertainty relation. In such procedure it is implicitly assumed that there is coherence in energy. Then the error in the above identification of the vector $\vec{f}_2$ as a vector in $M_3(1)$ is such that
\begin{align}
\widetilde{\Delta}=\,\tau\,\Delta E_G \sim \,\hbar,
 \end{align}
 for any quantum system, large or small. Therefore, according to new measure $\widetilde{\Delta}$, the amount of violation of the general covariance principle due to quantum coherence does not depend upon the size of the system, since it is always of order $\hbar$, despite that the lifetime $\tau$ could be large or small, depending on the size of the system.

The form of the paradox that we have reached by further pursuing Penrose's argument is the following. Although Penrose's argument is presented as an aim to preserve in an approximated way general covariance as much as it could be possible in settings where superpositions of gravitational spacetimes can be of relevance, the argument leads to a mechanism that violates general covariance in an observable way. Adopting Penrose's measure $\Delta$ and his explanation of the reduction of the wave function for macroscopic objects, then the same interpretation as in Penrose's theory yields to infer that the experimental observation of quantum coherence must be interpreted as an observable violation of general covariance. If one instead adopts the measure $\widetilde{\Delta}$, then there is no objective reason to attribute a small violation of the principle of general covariance for microscopic systems and large violation for large or macroscopic systems, because the violation of general covariance measured using $\widetilde{\Delta}$ is universal and the same for all systems obeying Heisenberg energy/time uncertainty relation.

\section{Potential resolutions of the conundrum and discussion}
One can cast doubts that the assumptions of Penrose's argument are valid. In particular, the identification of vector fields defined over different spacetimes is against common use in differential geometry. Hence this opens one obvious door to skip the consequences of Penrose's argument.

If the assumptions and methods used in Penrose's theory are tenable and an adequate framework for the manipulations required by Penrose's argument exists, then the paradox described in the previous section between the aims of the theory and the consequences of theory arises. There are two natural ways to resolve the paradox. One option is to preserve general covariance. Then a suppression of quantum coherence is expected, even for systems whose scales are considered to be at the quantum level. According to this point of view, it is impossible to attribute an ontological character to Heisenberg's energy/time uncertainty relation for small quantum systems when gravitational effects are taken into consideration. Obviously, this consequence is in conflict with the experimental evidence of quantum coherence at microscopic level. Therefore, to make consistent general covariance with quantum coherence, there must be an alternative interpretation of the quantum interference phenomenology to standard quantum mechanics able to accommodate these issues. Proposals of such style have been developed in \cite{Hooft2012wavefunctionschroedingercollapse, Hooft2016}  and in certain emergent quantum mechanics framework \cite{Ricardo2014, Ricardo2020b}. In such quantum emergent framework, gravity plays a fundamental role in the {\it reduction mechanism}, although it differs from the theories developed by Penrose \cite{Penrose 1996, Penrose 2005, Penrose 2014a} and by Di\'osi \cite{Diosi 1987, Diosi 1989} in a radical way. In particular, the theory presented in \cite{Ricardo2014,Ricardo2020b} is consistent with the principle of general covariance as it appears in general relativity. Furthermore, the gravitational interaction appears as a classical, emergent interaction and quantum coherence is seen not as an ontological process, but only as an epistemological characteristic of the effective quantum description.

The second possibility to resolve the paradox (assuming that Penrose's assumptions are valid) is to admit that, if the Heisenberg energy/time uncertainty relation is applicable, then the principle of general covariance as it is understand in general relativity is violated. This line of research is against the spirit of Penrose's argument, which was motivated as an attempt to preserve the principle of general covariance from being violated. However, if one is ready to accept the observability of the violation of general covariance, then other formulations of the mechanism of gravitational reduction are possible (\cite{Diosi 1987, Diosi 1989}. In this case, the gravitationally induced reduction of the wave function mechanism will provide an example of how the collapse can happen. However, note that for the class of theories where physical equations can be formulated in terms of geometric objects living in a spacetime manifold (geometric framework), there are theoretical reasons to preserve general covariance. In such settings, the accepted viewpoint is that general covariance can always be satisfied. Either a theory is make general covariant by hand, a procedure that is on the basis of Kretschmann argument \cite{Kreschtmann 1917}, or general covariance is a consistent requirement, as happens in general relativity for a consistent implementation of the equivalence principle in the mathematical form \cite{Bergmann 1976}, Chapter X, \cite{Penrose 2005}, section 19.6, or general covariance is useful as heuristic principle in the selection of physical models \cite{Einstein 1918}. 

We have discussed the original proposal of Penrose as presented in \cite{Penrose 1996}. However, it is necessary to say that Penrose has recently presented a different argument based upon the notion of different vacuum associated with the application of the equivalence principle in quantum interferometric systems \cite{Penrose 2014a}. The effect discussed is a non-relativistic remnant of {\it Davies-Unruh effect} \cite{Fulling 1973,Davies 1975,Unruh 1976}, since different relative accelerated observes describe the physical system with associated different vacuum states. The evolution of gravitational systems is also related with different vacuum, with different notions of positive frequencies. Such a confrontation between fundamental principles is, again, heuristically address by using Heinsenberg Energy/time uncertainty relation, that provides a mechanism for gravitational reduction of superpositions to stationary states. The proper understanding of how this new mechanism is affected by the argument presented in our paper requires further study.

Finally, let us mention that Penrose's theory has lead to falsifiable experimental proposals \cite{Howl et al.,Marshall et al.} and has also been, in conjunction with the work of Di\'osi \cite{Diosi 1987,Diosi 1989}, the driven force towards an experimental test of the so called Di\'osi-Penrose theory has been performed recently \cite{Donadi et al.}. Donadi et al. experiment presents evidence against the parameter-free version of the Di\'osi–Penrose model \cite{Diosi 1987, Diosi 1989}, although it does not fully rule out the general form of parameter-free models.
\section{Conclusion}
In conclusion, we have argued that under the assumptions and methods discussed by Penrose in \cite{Penrose 1996}  either leads to a renounce of the ontological applicability of the concept of coherence at the quantum level in the form of energy/time uncertainty relation, or leads to a re-interpretation of quantum coherence as an observable violation of general covariance.

\end{document}